\documentstyle[twoside,fleqn,espcrc2,epsf]{article}

\title{
The Scalar Quarkonium Spectrum and Quarkonium-Glueball Mixing}

\author{W.\ Lee \thanks{present address:
T-8, LANL, Los Alamos, NM 87545}
and D.\ Weingarten\\
IBM Research, P.O.~Box 218,
Yorktown Heights, NY 10598\\}

\begin{document}

\begin{abstract}

We evaluate the valence approximation to the mass of scalar quarkonium
and to the mixing energy between scalar quarkonium and the lightest
scalar glueball for a range of different lattice sizes and quark masses.
Our results support the identification of $f_0(1710)$ as the lightest
scalar glueball.
 
\end{abstract}

\maketitle

\section{Introduction}

The identification of $f_0(1710)$ as the lightest scalar glueball is
supported at present by two different sets of calculations.  A valence
approximation calculation on GF11~\cite{Sexton95} of the couplings for
the lightest scalar glueball to all possible pseudoscalar pairs, using a
$16^3 \times 24$ lattice at $\beta = 5.7$, gives $108 \pm 29$ MeV for
the total two-body width. Based on this number, any reasonable guesses
for the effect of finite lattice spacing, finite lattice volume, and the
remaining width to multibody states yield a total width small enough for
the lightest scalar glueball to be seen easily in experiment.  For the
valence approximation to the infinite volume continuum limit of the
lightest scalar glueball mass, a reanalysis~\cite{latestglue} of a
calculation on GF11~\cite{Vaccarino}, using 25000 to 30000 gauge
configurations, gives a mass of 1710(63) MeV.  An independent
calculation by the UKQCD-Wuppertal~\cite{Livertal} collaboration, using
1000 to 3000 gauge configurations, when extrapolated to the continuum
limit according to Ref.~\cite{Weingarten94} gives a mass of 1625(92)
MeV.  The two results combined give 1693(60) MeV.

Among established resonances with the quantum numbers to be a scalar
glueball, aside from $f_J(1710)$ all are clearly inconsistent with the
mass calculations expect $f_0(1500)$.  Refs.~\cite{Sexton95,Lee97}
interpret $f_0(1500)$ as dominantly composed of $s\overline{s}$ scalar
quarkonium.  A problem with this interpretation, however, is that
$f_0(1500)$ does not seem to decay mainly into states containing an $s$
and an $\overline{s}$ quark~\cite{Amsler1}. In part for this reason,
Ref.~\cite{Amsler2} interprets $f_0(1500)$ as the lightest scalar
glueball and $f_0(1710)$ as $s\overline{s}$ scalar quarkonium.  A second
problem is that the Hamiltonian of full QCD couples quarkonium and
glueballs so that $f_J(1710)$ and $f_0(1500)$ could both be linear
combinations of quarkonium and a glueball, perhaps even half glueball
and half quarkonium each.

For several different values of lattice spacing, lattice volume and
quark mass, we have now calculated the valence approximation to the mass
of the lightest scalar $q\overline{q}$ state and the mixing energy
between these states and the lightest scalar glueball. The infinite
volume continuum limit of the mass of the scalar $s\overline{s}$ state
we find lies at least 200 MeV below the mass of the lightest scalar
glueball, ruling out the interpretation of $f_0(1710)$ as an
$s\overline{s}$ state with $f_0(1500)$ as a glueball. For the mixing
energy evalulated at the strange quark mass, $E(m_s)$, and at the average
light quark mass, $E((m_u + m_d)/2)$, we find the infinite volume
continuum limit of $E((m_u + m_d)/2)/E(m_s)$ to be 1.31(12).  If this
ratio is placed in the model of Ref.~\cite{Weingarten97} we find that
$f_0(1500)$ is at least 75\% $s\overline{s}$ quarkonium and that
$f_0(1710)$ is at least 75\% glueball.  In addition, we find that
$f_0(1500)$ acquires a significant component of $(u\overline{u} +
d\overline{d})/\sqrt{2}$ from mixing with $f_0(1390)$, composed mainly
of $(u\overline{u} + d\overline{d})/\sqrt{2}$. The $s\overline{s}$ and
$(u\overline{u} + d\overline{d})/\sqrt{2}$ components of $f_0(1500)$,
however, have opposite sign, interfere destructively in decay to
$K\overline{K}$ final states and suppress the partial width of
$f_0(1500)$ by a factor of $0.5$ or less in comparison to an unmixed
$s\overline{s}$ state. This suppression provides an explanation for the
small observed $K\overline{K}$ partial width.

Our calculations were done on ensembles of 2288 configurations on $ 16^2
\times 14 \times 20$ at $\beta$ of 5.93, 1781 configurations on $24^4$
at $\beta$ of 5.93, 582 configurations on $24^2 \times 20 \times 32$ at
$\beta$ of 6.17, and 1003 configurations on $32^2 \times 28 \times 40$
at $\beta$ of 6.40.  To vacilitate extrapolation to zero lattice
spacing, the smaller lattice with $\beta$ of 5.93, and the lattices with
$\beta$ of 6.17 and 6.40 were chosen to be nearly the same size in units
of the rho Compton wavelength, with periods $m_{\rho} L$ of 6.1(1),
6.6(2) and 6.3(2), respectively~\cite{Butler}, in the two equal space
directions.

\begin{figure}
\epsfxsize=63mm
\epsfbox{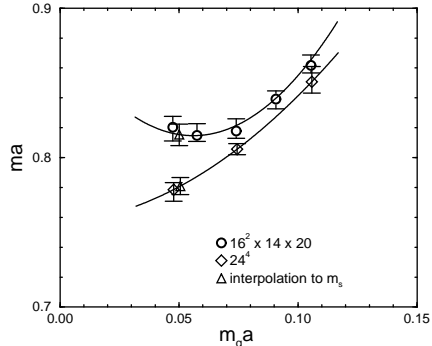}
\vskip -14mm
\caption{Scalar quarkonium mass as a function of quark mass for
$\beta$ of 5.93}
\vskip -10mm
\label{fig:mass}
\end{figure}

\begin{figure}
\epsfxsize=63mm
\epsfbox{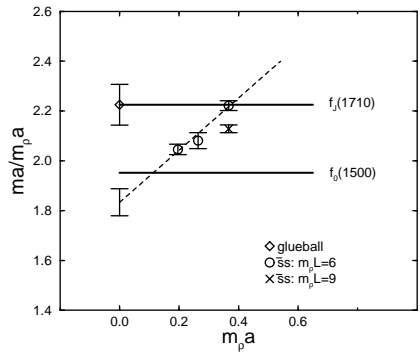}
\vskip -14mm
\caption{Continuum limit of the scalar $s\overline{s}$ mass.}
\vskip -7mm
\label{fig:masscont}
\end{figure}

Following Ref~\cite{Lee97} we evaluated on each ensemble correlation
functions with random sources built from quark and antiquark fields.
Averaging over random sources yields the correlation functions
\begin{eqnarray}
\label{defC}
C_{ff}(t) & = & \sum_{\vec{x}} < f(\vec{x},t) f(0,0)>, \nonumber \\
C_{g\sigma}(t) & = & \sum_{\vec{x}} < g(\vec{x},t) \sigma(0,0)>, 
\end{eqnarray}
for $f$ either $\sigma$, $\pi$ or $g$,
where $\sigma(\vec{x},t)$ and $\pi(\vec{x},t)$ are, respectively, the smeared
scalar and pseudoscalar operators of Ref.~\cite{Lee97} built from a
quark and an antiquark field and $g(\vec{x},t)$ is the smeared scalar
glueball operator of Ref.~\cite{Vaccarino}. We omit, for lack of space,
a discussion of our choices of smearing parameters.

From the large $t$ asymptotic
behavior of the diagonal correlators, 
\begin{eqnarray}
\label{diagZm}
C_{ff}(t) & \rightarrow & Z_f exp(-m_f t),
\end{eqnarray}
we found the masses $m_{\sigma}$, $m_{\pi}$, 
and $m_g$ and field strength renormalization constants $Z_{\sigma}$, $Z_{\pi}$ and
$Z_g$. From the asymptotic behavior of the off-diagonal correlator for
$m_{\sigma}$ close to $m_g$
\begin{eqnarray}
\label{offdiag}
\lefteqn{C_{g\sigma}(t) \rightarrow } \\
& & \sqrt{Z_g Z_{\sigma}} E \nonumber  
\sum_{t'} exp( -m_{\sigma} |t - t'| - m_g | t'|) \nonumber
\end{eqnarray}
we then found the glueball scalar-quarkonium mixing energy E.
Eqs.~(\ref{diagZm}) and (\ref{offdiag}) have been simplified by omitting
terms arising from propagation around the lattice's periodic time
boundary.

Figure~\ref{fig:mass} shows the scalar quarkonium mass as a function of
quark mass for the two lattices with $\beta$ of 5.93. For both lattices
the solid lines are fits to quadratic functions of quark mass which we
used to interpolate to the strange quark mass. The strange quark mass we
chose as the value which yields, in physical units, a pseudoscalar mass
squared of $2 m_K^2 - m_{\pi}^2$, for $m_K$ and $m_{\pi}$ the observed
kaon and pion masses, respectively. As shown by the figure, for the
lattice $16^2 \times 14 \times 20$ with $m_{\rho} L$ of $6.1(1)$ the
scalar mass as a function of quark mass flattens out as quark mass is
lowered toward the strange quark mass and then actually begins to rise
as the quark mass is decreased still further. This feature is absent
from the data for $24^4$ with $m_{\rho} L$ of $9.2(2)$ at $\beta$ of
5.93 and is thus a finite-volume artifact. It is present also in the
data, not shown, at $\beta$ of 6.17 with $m_{\rho} L$ of 6.6(2) and at
$\beta$ of 6.40 with $m_{\rho} L$ of 6.3(2).

Figure~\ref{fig:masscont} shows a linear extrapolation to zero lattice
spacing of the $s\overline{s}$ scalar mass, measured in units of the rho
mass, for $m_{\rho} L$ near 6, along with the point at $\beta$ of 5.93
with $m_{\rho} L$ near 9. It is clear that for $m_{\rho} L$ of 6, the
continuum limit of the $s\overline{s}$ scalar mass lies more than 200
MeV below the valence approximation to the scalar glueball mass. 
Figure~\ref{fig:masscont} shows also
that as $m_{\rho} L$ is increased from 6 to 9, the
$s\overline{s}$ scalar mass falls. Thus the infinite volume
continuum limit of the $s\overline{s}$ scalar mass should be more than
200 MeV below the scalar glueball mass.

\begin{figure}
\epsfxsize=63mm
\epsfbox{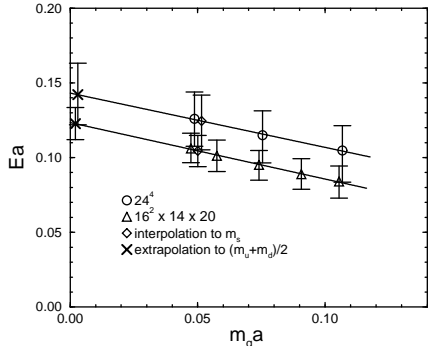}
\vskip -14mm
\caption{Glueball-quarkonium mixing energy as a function of
quark mass for $\beta$ of 5.93}
\vskip -12mm
\label{fig:mix}
\end{figure}

\begin{figure}
\epsfxsize=63mm
\epsfbox{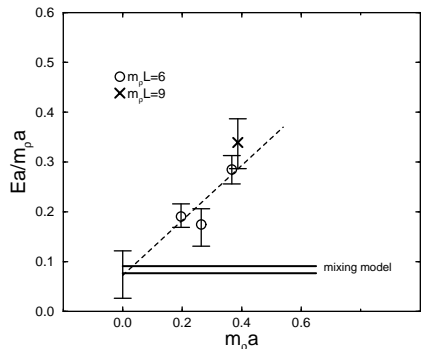}
\vskip -14mm
\caption{Continuum limit of the glueball-quarkonium mixing energy
with $m_{\rho} L$ of 6.}
\vskip -9mm
\label{fig:mixcont}
\end{figure}

Figure~\ref{fig:mix} is the analogue of Figure~\ref{fig:mass} for
quarkonium-glueball mixing energy. For mixing energy the quark mass
dependence shows no anomoly. The mixing energies at difference quark
masses turn out to be highly correlated and depend quite linearly on
quark mass. For $\beta$ of 6.17 and 6.40 the mixing energy behaves
similarly. It appears that the mixing energy can be extrapolated
reliably down to the average light quark mass $(m_u + m_d)/2$. For the
data at $\beta$ of 5.93, the mixing energy ratio $E((m_u +
m_d)/2)/E(m_s)$ is 1.17(3) for $m_{\rho} L$ of 6.1(1) and 1.14(5) for
$m_{\rho} L$ of 9.2(2). The ratio has at most rather small volume
dependence and thus seems already to be near its infinite volume limit
with $m_{\rho} L$ of 6.1(1).

Figure~\ref{fig:mixcont} shows a linear extrapolation to zero lattice
spacing of quarkonium-glueball mixing energy at the strange quark mass
giving a limit of of 56(37) MeV.  A linear extrapolation of the ratio
$E((m_u + m_d)/2)/E(m_s)$ gives a continuum limit of 1.31(12).

If our predicted value of the ratio of mixing energies is used in the
mixing model of Ref.~\cite{Weingarten97}, we obtain the mixing results
described earlier. With this choice of mixing energy ratio, an
additional predicition of the model is that the state $f_0(1400)$,
composed mainly of $(u\overline{u} + d\overline{d})/\sqrt{2}$, picks up
most of the glueball amplitude that leaks out of $f_0(1710)$ and
therefore should be significantly more prominent in $J/\Psi$ radiative
decays than is the $f_0(1500)$. This prediction is consistent with SLAC
Mark III data~\cite{MarkIII}.  For the mixing energy evaluated at the
strange quark mass, $E(m_s)$, the model predicts 72(6) MeV.  The one
sigma upper and lower bounds on the model's prediction are shown as
horizontal lines in Figure~\ref{fig:mixcont} and are consistent with the
extrapolated continuum value.

\end{document}